\PassOptionsToPackage{dvipsnames}{xcolor}
\documentclass[times]{aastex631}

\usepackage{amsmath}
\usepackage{CJK}
\usepackage{bm}
\usepackage[normalem]{ulem}
\usepackage{enumitem}

\makeatletter
\newcommand*{\rom}[1]{\expandafter\@slowromancap\romannumeral #1@}
\makeatother

\newcommand{\jl}[1]{\textcolor{black}{#1}}

\usepackage{newunicodechar,graphicx}
\DeclareRobustCommand{\okina}{%
  \raisebox{\dimexpr\fontcharht\font`A-\height}{%
    \scalebox{0.8}{`}%
  }%
}

\begin{document}

\begin{CJK*}{UTF8}{gbsn}
\title{Large Photospheric Doppler Shift in Solar Active Region 12673: \rom{1}. Field-Aligned Flows}

\author[0000-0002-7290-0863]{Jiayi Liu (刘嘉奕)}
\affiliation{Institute for Astronomy, University of Hawai$\okina$i at M\={a}noa, 2680 Woodlawn Dr., Honolulu, HI 96822, USA}

\author[0000-0003-4043-616X]{Xudong Sun (孙旭东)}
\affil{Institute for Astronomy, University of Hawai$\okina$i at M\={a}noa, 34 Ohia Ku Street, Pukalani, HI 96768, USA}

\author[0000-0003-1522-4632]{Peter W. Schuck}
\affil{Space Weather Laboratory, NASA Goddard Space Flight Center, Greenbelt, Maryland, US}

\author[0000-0001-5459-2628]{Sarah A. Jaeggli}
\affil{National Solar Observatory, 22 Ohia Ku Street, Pukalani, HI 96768, USA}

\author[0000-0003-2244-641X]{Brian T. Welsch}
\affil{Natural and Applied Sciences, UW-Green Bay, 2420 Nicolet Drive, Green Bay, WI, USA}

\author[0000-0001-5518-8782]{Carlos Quintero Noda}
\affil{Instituto de Astrof\'{i}sica de Canarias, E-38205 La Laguna, Tenerife, Spain}
\affil{Departamento de Astrof\'{i}sica, Universidad de La Laguna, E-38206 La Laguna, Tenerife, Spain}

\correspondingauthor{Xudong Sun (孙旭东)}
\email{xudongs@hawaii.edu}


\begin{abstract}
Delta ($\delta$) sunspots sometimes host fast photospheric flows along the central magnetic polarity inversion line (PIL). Here we study the strong Doppler shift signature in the \jl{central penumbral} light bridge of solar active region \jl{NOAA 12673}. Observations from the Helioseismic and Magnetic Imager (HMI) indicate highly sheared, strong magnetic fields. Large Doppler shifts up to 3.2~km~s$^{-1}$ appeared during the formation of the light bridge and persisted for about 16 hours. A new velocity estimator, called DAVE4VMwDV, reveals fast converging and shearing motion along the PIL from HMI \jl{vector magnetograms}, and recovers the observed Doppler signal much better than an old version of the algorithm. The inferred velocity vectors are largely (anti-)parallel to the inclined magnetic fields, suggesting that the observed Doppler shift contains significant contribution from the projected, field-aligned flows. High-resolution observations from the \jl{Hinode}/Spectro-Polarimeter (SP) further exhibit a clear correlation between the Doppler velocity and the cosine of the magnetic inclination, which is in agreement with \jl{HMI results} and consistent with a field-aligned flow of about $9.6$~km~s$^{-1}$. The complex Stokes profiles suggest \jl{significant gradients} of physical variables along the line of sight. We discuss the implications on the $\delta$-spot magnetic structure and the flow-driving mechanism.
\end{abstract}

\keywords{\href{http://astrothesaurus.org/uat/1975}{Solar active region magnetic fields (1975)}; \href{http://astrothesaurus.org/uat/1976}{Solar active region velocity fields (1976)}; \href{http://astrothesaurus.org/uat/1979}{Delta sunspots (1979)}}

\vspace{6mm}


\section{Introduction} \label{sec:intro}

\subsection{Delta Sunspots and Photospheric Doppler Shift}

A significant fraction of solar flares and coronal mass ejections (CMEs) \jl{originates} from active regions. Complex active regions with ``delta ($\delta$) type'' sunspots \jl{\citep{Hale, kunzel}}, in which the umbrae of positive and negative polarities share a common penumbra, tend to be the most flare-productive \citep{flare}. Statistical studies show that more than 80\% of the \jl{GOES} X-class flares occur in $\delta$-spots with strong magnetic fields \citep{flare_stat_2000,flare_stat_2014}. 

\jl{Flares and CMEs originate from above the polarity inversion line (PIL), where the photospheric magnetic field changes sign.} The horizontal photospheric field tangent to the PIL represents the ``sheared'' component of the total horizontal field. It often correlates with enhanced electric current density and magnetic free energy, which are required to drive eruptions. As expected, the magnetic fields along the PILs of $\delta$-spots are often strong and highly sheared. \cite{okamoto} observed that the strength of magnetic fields can reach 6250~G in the PIL. \citet{shear} reported that the magnetic shear at a flaring PIL, defined as the angle between the observed and the potential field, can be as high as $80^{\circ}$ to $90^{\circ}$.

While convection in the dark umbrae is mostly suppressed, photospheric plasma motions along the $\delta$-spot PILs can be highly dynamic. Many observers noticed long-lasting, large Doppler velocities within the \jl{central penumbral} light bridge or penumbral filaments. Spectropolarimetric studies of some $\delta$-spots have shown Doppler shifts ranging from a few \jl{kilometers per second} \citep{fast_flow_takizawa,fast_flow_shimizu,fast_flow_2016} to supersonic speeds \citep[14~km~s$^{-1}$;][]{fast_flow_1994}.


\begin{figure*}[t!]
\centering\includegraphics[width=\textwidth]{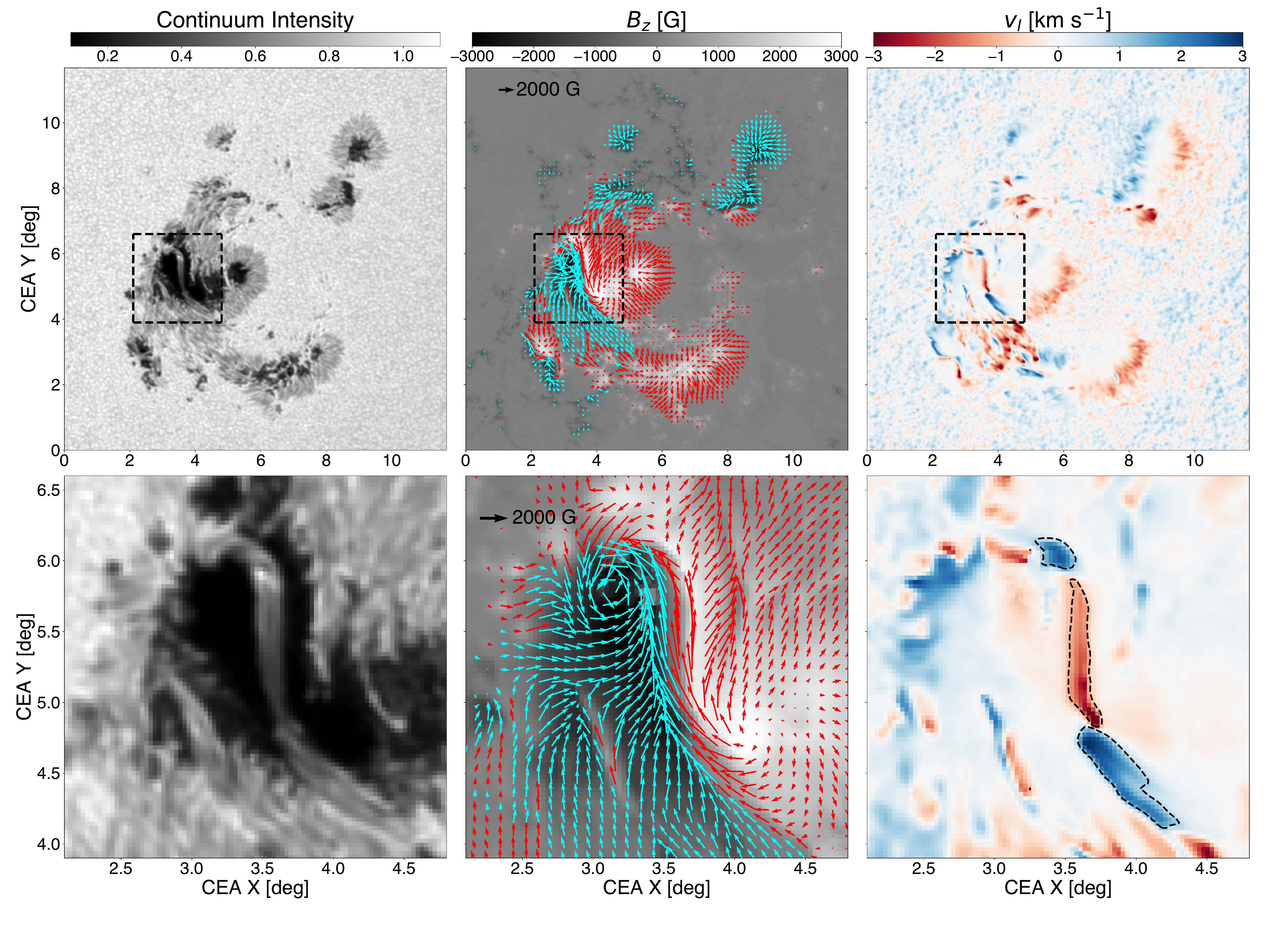}

\caption{Overview of active region NOAA 12673 from HMI observation at September 6 00:00 UT. The top row shows the entire active region, while the bottom row zooms in on the central region, within the dashed box in each top-row panel, which contains the \jl{central penumbral} light bridge and large Doppler shift signal of interest. Left: continuum intensity normalized to nearby quiet-Sun mean. Middle: vector magnetic field map. The background gray image shows the vertical field. The red (cyan) arrows shows the horizontal field vectors in positive- (negative-) $B_z$ regions. The scale of field strength is plotted on the map. Right: Doppler velocity after various corrections (see Section~\ref{subsec:hmi_method}). Redshifted (blueshifted) regions have negative (positive) \jl{LOS} velocity. Note that opposite to the convention, the sign of $v_l$ is positive (negative) if the flow is toward (away from) observer. The dashed contours are for $v_l=\pm1$~km~s$^{-1}$. The dotted boxes in the top row indicate the field of view of the bottom row.}
\vspace{2mm}
\label{fig:roi}
\end{figure*}


The exact nature of these flows remains unclear. On the one hand, fast magnetic flux emergence or flux cancellation may induce Doppler signals. On the other hand, some have interpreted the observations as the projection of field-aligned flows such as the Evershed flow \citep{fast_flow_Lite,fast_flow_2016}. As the interaction between strong magnetic field and fast flows can bring significant magnetic energy and helicity into the corona to enhance the \jl{active region} eruptive potential \citep{welsch2009}, the phenomenon warrants further investigation.


\subsection{Overview}

Active region NOAA 12673 is the most flare-productive region of solar cycle 24. It produced four X-class flares during its passage across the solar disk in early September 2017 \citep{12673_flare}. Among them, the X9.3-class flare (SOL2017-09-06T11:53) was the most intense flare since 2005.

\jl{This active region is well studied including its magnetic fields and horizontal flow fields.} It started as a decaying region containing a single sunspot of positive polarity. Multiple pairs of bipoles successively emerged to the east within a few days. According to \cite{sun2017}, it is one of the fastest emergence events ever reported. The interaction between these bipoles resulted in a complex active region. A strong-field, highly-sheared PIL gradually formed in the core region along with a narrow \jl{central penumbra} (Figure~\ref{fig:roi}). \cite{Wang_2018} reported an extremely strong field strength that reached 5570~G in the northern portion late on September 6. \jl{The study of horizontal flows by \cite{12673flow} suggested that the persistent shear motion near the PIL contributed to the highly non-potential coronal field, which provided the energy for the X-class flares.}

\jl{This paper aims to investigate the properties and the origin of the large photospheric Doppler shift observed in active region NOAA 12673. Specifically, we infer the magnitude and direction of the flows in the central light bridge using a recently developed flow-tracking method and vector magnetograms from the Helioseismic and Magnetic Imager \citep[HMI;][]{schou2012}. We further study the fine-scale, height-dependent flow and magnetic field using high resolution observations from the Spectro-Polarimeter \citep[SP;][]{Lite2013}.} The signal appeared \jl{at about 18:00 UT on September 5} as large redshift in the \jl{central penumbra}, which persisted for about 16 hours (Figure~\ref{fig:roi}). It gradually weakened and almost disappeared before the X9-class flare on September 6. Here we focus on a short period around \jl{00:00 UT, September 6} when the Doppler signal was fully developed for detailed study. The active region centroid was then located at about $30\degr$ west and $9\degr$ south in the heliographic coordinate frame.

We describe the observational data and the analysis methods in Section~\ref{sec:D_M}. We present the results in Section~\ref{sec:results}. We discuss caveats of the study and the possible flow-driving mechanism in Section \ref{sec:Discussion}, and finally conclude in Section~\ref{sec:conclusion}.


\section{Observation and Data Analysis} \label{sec:D_M}

\subsection{SDO/HMI Observation and Velocity Inference}\label{subsec:hmi_method}

The Helioseismic and Magnetic Imager onboard the \jl{Solar Dynamics Observatory} \citep[SDO;][]{SDO} measures the full-disk Stokes profiles at six wavelengths across the photospheric \ion{Fe}{1} 617.3~nm absorption line. These data are used to infer the magnetic vector $\bm{B}$ and line-of-sight (LOS) velocity $v_l$ in the photosphere with a 12~minute cadence and a $0\farcs5$ plate scale. 

In this study, we use the \jl{vector magnetogram} data provided by the Space Weather HMI Active Region Patch \citep[SHARP;][]{bobra2014} dataset, which automatically tracks strong concentrations of photospheric magnetic field across the solar disk. With the active region center tracked at a pre-determined differential rotation, we produce $\bm{B}$ and $v_l$ maps using a cylindrical equal-area \jl{(CEA)} projection \citep{sun2013arXiv}. These maps have a pixel size of $0.03^\circ$, equivalent to about 360~km. We further remove from the Doppler maps the contribution of differential rotation, the relative movement of SDO with respect to the Sun, and the convective blueshift bias \citep{welsch2013}.

While the LOS velocity $v_l$ can be estimated from the Doppler effect, there is no direct observational access to the full vector velocity field $\bm{v}$. Various algorithms have been developed to infer the photospheric flow fields \citep[e.g.,][]{welsch2007,dave}. \jl{For example, the algorithms based on local correlation tracking techniques \citep[LCT; e.g.,][]{Hurlburt1995,FLCT} can be used to determine the local displacement of features between two input successive images. These techniques, however, are only sensitive to advection based on the change of spatial patterns, and can miss important terms that contribute to the magnetic energy and helicity injection (such as rotation of an axisymmetric sunspot). The performance of velocity inversion algorithms can be improved by taking advantage of the physical relation between $\bm{v}$ and $\bm{B}$ \citep[e.g.,][]{Kusano2002,welsch2004, Longcope2004, schuck2006, dave}, which takes the form of the normal component of ideal magnetic induction equation:
\begin{equation}\label{equ:ideal}
    \frac{\partial{B_z}}{\partial t} = - \nabla_h \cdot (B_z\bm{v}_h+\bm{B}_hv_z).
\end{equation}
One may infer $\bm{v}$ from a time sequence of $\bm{B}$ maps using Equation~(\ref{equ:ideal}). Due to the lack of information in the evolution of horizontal magnetic field, the inversion problem is not well-posed and generally requires additional assumptions or constraints.}

The Differential Affine Velocity Estimator for Vector Magnetograms \citep[DAVE4VM;][]{dave} is a widely used, local optical flow velocity estimator. Within a windowed subregion, it attempts to minimize the L2 norm of the normal component of Equation~(\ref{equ:ideal}). Recently, we modified the algorithm to include a more complex velocity model and the observed Doppler velocity $v_l$ as an additional constraint, which we dub DAVE4VM with Doppler Velocity (DAVE4VMwDV; Schuck, in prep.).

The loss function $L$ of DAVE4VMwDV reads
\begin{equation}
\begin{split}
L &= L_1 + \lambda L_2, \\ 
L_1 &= \sum_w \left\{ \frac{\partial B_z}{\partial t} + \bm\nabla _h \cdot (B_z \bm{v}_h+\bm{B}_hv_z)\right\}^2 \sigma^{-2}_{\partial_t B_z}, \\
L_2 &= \sum_w (\hat{\bm{\eta}}\cdot \bm{v}-v_l)^2 \sigma^{-2}_{v_l}.
\label{l2}
\end{split}
\end{equation}
where $\bm{B}$ and $v_l$ are the observational input, and $\bm{v}$ is the output. The subscripts $z$ and $h$ indicate the normal and horizontal components respectively; $\bm{\nabla}_h$ acts on the horizontal components alone. The unit vector $\hat{\bm{\eta}}$ specifies the LOS direction. \jl{In practice, we use the difference of two successive frames divided by the time cadence $\Delta t$ to calculate the time derivative, and a five-point stencil for the spatial derivatives:
\begin{equation}
  \begin{split}
  \frac{\partial f_{i,j}}{\partial t} &= \frac{f_{i,j}(t+\Delta t) - f_{i,j}(t)}{\Delta t}, \\ 
  \frac{\partial f_{i,j}}{\partial x} &= \frac{-f_{i+2,j}+8f_{i+1,j}-8f_{i-1,j}+f_{i-2,j}}{12}, \\
  \frac{\partial f_{i,j}}{\partial y } &= \frac{-f_{i,j+2}+8f_{i,j+1}-8f_{i,j-1}+f_{i,j-2}}{12},
  \label{diff}
  \end{split}
  \end{equation}
}
\jl{where $i$ and $j$ represent the index of the grid points in the $x$ and $y$ direction, respectively.}
For this study, we assume that the HMI CEA maps sample $\bm{B}$ and $v_l$ on a uniform local Cartesian grid at a constant geometric height.

We note that in this work, the positive (negative) $v_l$ denotes blueshifted (redshifted) regions such that it is consistent with the sign of $v_z$ at disk center. This is opposite to the observer's convention of $v_l$. We nevertheless use blue (red) colors for the blueshifted (redshifted) regions in all figures (e.g., right column of Figure~\ref{fig:roi}).

In Equation~(\ref{l2}), the term $L_1$ is the original, DAVE4VM loss function that describes the residual of the vertical component of the induction equation. The term $L_2$ is the new Doppler constraint that penalizes the differences between the observed and the inferred LOS velocity, whose relative importance is controlled by a free parameter $\lambda$. The sum is performed over all pixels in each window of size $w$. We note that $L_1$ and $L_2$ are normalized by the respective uncertainties in the quantities derived from observations, $\sigma_{\partial_t B_z}$ and $\sigma_{v_l}$, respectively.

We use the following values of for the free parameters in DAVE4VMwDV: the window size for optimization is $w=23$~pixels; the maximum degree of Legendre polynomials $d=7$ (for velocity expansion inside the window), and the relative weighting $\lambda=0.5$. Appendix \ref{app:DAVE4VMwDV} provides more details on these free parameters.


\begin{deluxetable}{cccc}[t!]
\tablecaption{Summary of SIR algorithm
configuration}\label{tab:SIR}
\tablewidth{0pt}
\tablehead{
\colhead{}&
  \multicolumn{3}{c}{{Nodes}} \\
\cline{2-4}
\colhead{\textbf{Parameters}} & \colhead{\textbf{Cycle 1}} & \colhead{\textbf{Cycle 2}} & \colhead{\textbf{Cycle 3}} 
}
\startdata
  Temperature     & 2 & 3 & 5 \\ 
  Microturbulence & 1 & 1 & 1 \\ 
  LOS velocity    & 1 & 2 & 3 \\ 
  Magnetic field strength  & 1 & 2 & 3 \\ 
  Inclination     & 1 & 2 & 3 \\ 
  Azimuth         & 1 & 1 & 2 \\ 
\enddata
\end{deluxetable}
\vspace{-5mm}


\begin{figure*}[t!]
\centering\includegraphics[width=\textwidth]{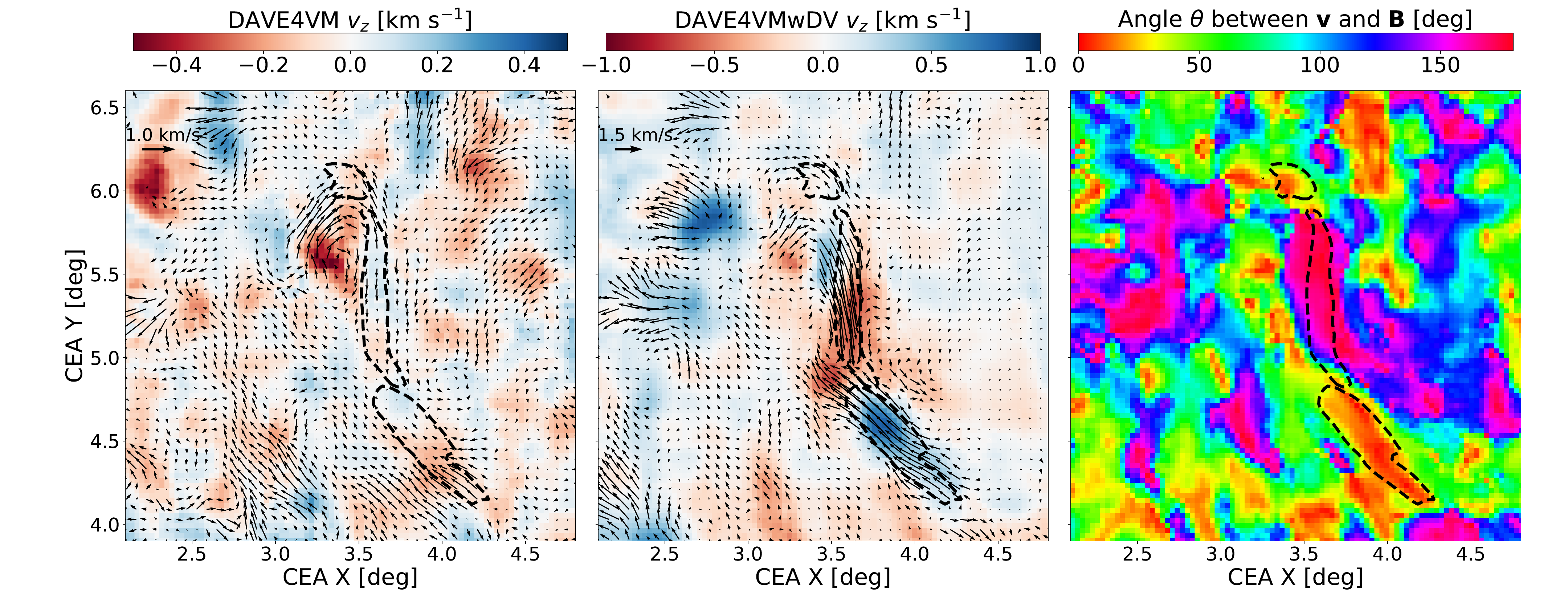}
\caption{Photospheric velocity field in the active region core region based on HMI data. Left: map of inferred $\bm{v}$ from DAVE4VM, shown as a comparison to the nominal result in the next panel. The background image shows the \textit{vertical} velocity $v_z$. The arrows illustrate the horizontal velocity $\bm{v}_h$. Middle: similar to left, but from DAVE4VMwDV. Right: map of the angle $\theta$ between $\bm{v}$ and $\bm{B}$. Red color indicates the region with flows parallel or anti-parallel to the magnetic field. The dashed contours are for \textit{observed} Doppler velocity $v_l=\pm1$~km~s$^{-1}$, defined in the last panel of Figure~\ref{fig:roi}.}
\label{fig:flow}
\vspace{2mm}
\end{figure*}


\subsection{Hinode/SP Observation and Stokes Inversion}

The Spectro-Polarimeter on board the \jl{Hinode} satellite \citep{Hinode_satellite} carries out high-resolution ($0\farcs3$) spectropolarimetric observations of the Sun's photosphere. \jl{Spectro-Polarimeter} is a slit-spectrograph instrument. It measures the four Stokes parameters of two magnetically sensitive \ion{Fe}{1} lines at 630.15 and 630.25 nm with a 21.5 m\AA~spectral sampling rate. The wide spectral window and fine spectral resolution allow for the study of depth-dependent structure.

In this study, we analyze several SP level-1 Stokes rasters of active region NOAA 12673. We focus on the observations starting at 00:02 UT on Sep 6. The scan was performed in the ``normal mode'' with a spatial sampling of $0\farcs15$; the integration time per slit position is $4.8$~s. With 1024 scan locations and 1021 is pixels along the slit, the $1024\times1021$ ($151\arcsec \times 162\arcsec$) map took about 90 minutes to finish. Using the polarization signals in continuum, we estimate a noise level of $\sigma_Q \simeq \sigma_U \simeq \sigma_V \simeq 1.8\times 10^{-3} I_c$ for Stokes $Q$, $U$, and $V$, where $I_c$ is the continuum intensity.

We use the algorithm Stokes Inversion based on Response functions \citep[SIR;][]{SIR} to infer depth-dependent physical properties from SP Stokes data. SIR computes synthetic Stokes profiles by solving the radiative transfer equations for polarized light in a model atmosphere. In the inversion mode, SIR modifies the model atmosphere iteratively until the synthetic Stokes profiles match the observed ones. It finally returns temperature $T$, LOS velocity $v_l$, magnetic field strength $B$, inclination $\gamma$, and azimuth $\psi$ as a function of the optical depth $\tau_{500}$ (for continuum at 500~nm). 

During each iteration of the inversion, perturbations of the physical variable are introduced at specific locations along the LOS known as the ``nodes.'' The number of nodes is a free parameter; more nodes allow for greater variations along the LOS. In this study, we use a typical configuration listed in Table~\ref{tab:SIR}. We start with constant magnetic field and LOS velocity and only linear perturbations for temperature in the first cycle. We then gradually introduce more nodes to magnetic field, LOS velocity and temperature. Finally we use five nodes for temperature, three nodes for LOS velocity, magnetic field strength and inclination and two nodes for azimuth. \jl{Micro-turbulence in our model is height independent, so we use only one node.} For comparison, we also performed an inversion with only one node for all parameters, which has no gradients along the LOS.

Our inversion scheme \jl{takes into account} the spectral PSF provided in \cite{Lite2013} so we do not use any macro-turbulence correction. It also assumes a fixed magnetic filling factor of unity, which works reasonably well in the sunspot regions and \jl{is} consistent with the HMI inversion. Effects arising from a multiple-component atmosphere will be the focus of a future paper.


\begin{figure*}[t!]
\centering\includegraphics[width=\textwidth]{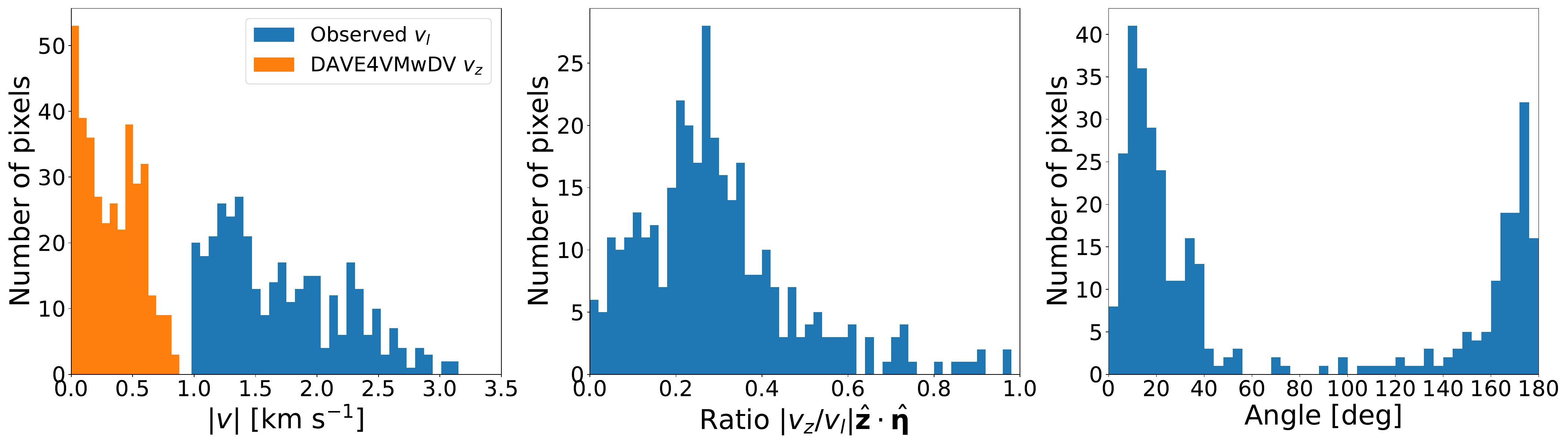}
\caption{Histogram of several variables in the central penumbral light bridge with fast Doppler flow (inferred $\lvert v_l \rvert \ge 1$~km~s$^{-1}$). The regions are marked by the contours in Figure~\ref{fig:flow}. Left: observed Doppler velocity $v_l$ (average of 00:00 and 00:12 UT frames) and inferred vertical velocity $v_z$.  Middle: variable $\lvert v_z / v_l \, \rvert \hat{\bm{z}} \cdot \hat{\bm{\eta}}$, where $\hat{\bm{z}}$ is the local vertical unit vector. 
This illustrates
the relative contribution of $v_z$ to the observed $v_l$. Right: angle between $\bm{v}$ and $\bm{B}$.}
\label{fig:histogram}
\vspace{2mm}
\end{figure*}


\section{Results}\label{sec:results}

\subsection{Photospheric Magnetic and Velocity Fields}\label{subsec:hmi}

Several features are noteworthy in HMI observations (Figure~\ref{fig:roi}). First, the \jl{central penumbra} takes the form of a narrow, inverse-S-shaped light bridge. The median width of the light bridge is about 2.2~Mm, and the median continuum intensity is 36\% of the nearby quiet Sun. There are hints of substructures across the light bridge, which are not well resolved at HMI's resolution. Second, the magnetic fields inside the \jl{central penumbra} are strong, highly inclined, and run almost parallel to the PIL. The median field strength is about 3.5~kG, the median cosine of the inclination angle (with respect to the local normal) is $-0.1$ (equivalent to $95\fdg7$), and the median shear angle is $75.9 \degr$.  Near the PIL, the horizontal field vectors display clear, counterclockwise and clockwise rotational patterns in the northern and southern umbrae, respectively. According to \cite{Schuck2022}, this pattern is unambiguously associated with electric currents through the photosphere. Third, the redshift signal is co-located with the \jl{central penumbra} with a median $v_l$ of $-0.8$~km~s$^{-1}$ and a maximum of $-3.2$~km~s$^{-1}$. Large blueshift signals that bracket the redshift from the north and the south have similar magnitude.

We use two HMI magnetograms from \jl{00:00 UT and 00:12 UT on September 6}, and the averaged Dopplergram from these two times to estimate the photospheric velocity field. As shown in the middle panel of Figure \ref{fig:flow}, the \textit{vertical} velocity $v_z$ (perpendicular to the surface) inferred from DAVE4VMwDV at the northern half of the central light bridge is mostly negative. This region is approximately co-spatial with the observed, redshifted region where $v_l\le-1$~km~s$^{-1}$ (dashed contour). Similarly, the elongated, positive $v_z$ patch in the south approximately coincides with the blueshifted region where $v_l\ge1$~km~s$^{-1}$. In contrast, $v_z$ is small in the northern end of the light bridge.

These results suggest that vertical up/downflow can only explain a portion of the observed Doppler signal. As shown in the left panel of Figure~\ref{fig:histogram}, the inferred $|v_z|$ is significantly smaller than the observed $|v_l|$ in regions where $|v_l|\ge1$~km~s$^{-1}$. \jl{To assess the contribution} of $v_z$ to $v_l$, we calculate the quantity $\lvert v_z / v_l \, \rvert \hat{\bm{z}} \cdot \hat{\bm{\eta}}$, where $v_z$ and $v_l$ are the inferred vertical and Doppler velocity from DAVE4VMwDV and $\hat{\bm{z}}$ is the local vertical unit vector. The middle panel of Figure~\ref{fig:histogram} shows \jl{a distribution peaking at $0.27$}.

The middle panel of Figure~\ref{fig:flow} also shows that the inferred horizontal velocity $\bm{v}_h$ is largely parallel to the PIL in the light bridge. The vectors display clear converging patterns around the southern end where $v_l$ changes sign. The vectors in the redshift region mostly point southward, while those in the blueshift regions point toward northeast. The magnitude can reach about 3~km~s$^{-1}$. We discuss in Section~\ref{subsec:floworigin} a possible magnetic geometry that could create such a flow pattern, as proposed by \citet{fast_flow_Lite}.

In addition to the converging flow, there also appears to be clockwise and counterclockwise rotational flows in the northern and southern umbrae, respectively. This is oppositely directed with respect to the magnetic field (Figure~\ref{fig:roi}), which may reduce the twist of the coronal field. 

We calculate the acute angle between the magnetic field and velocity vectors and show the result in the right panel of Figure~\ref{fig:flow}. The orange, red, and magenta colors occupy most of the contoured region, suggesting that the flows are largely (anti-)parallel to the magnetic field. Indeed, \jl{the} histogram of the angle in the right panel of Figure~\ref{fig:histogram} displays two striking peaks at about $20^{\circ}$ and $170^{\circ}$. We conclude that the Doppler signal contains significant contribution from the projected, \textit{field-aligned} flow.

The flows inferred from DAVE4VM (without Doppler constraint), shown in the left panel of Figure~\ref{fig:flow}, are quite different from DAVE4VMwDV. We discuss the implications in Section~\ref{subsec:flowtracking}.

The result is perhaps unsurprising given that the active region center is situated at about $34\degr$ away from the disk center ($\mu \simeq 0.85$). At such a heliocentric angle, the projection effect is starting to become significant. That is, the LOS velocity will have a non-negligible horizontal component. This is readily illustrated by the Doppler signals in the active region's outer penumbrae (Figure~\ref{fig:roi}), which come from the projected velocity of the Evershed flows. We have additionally derived the velocity maps for one hour before and after \jl{00:00 UT on September 6}. The velocity patterns remain similar during these two hours.


\begin{figure*}[t!]
  \centering\includegraphics[width=0.75\textwidth]{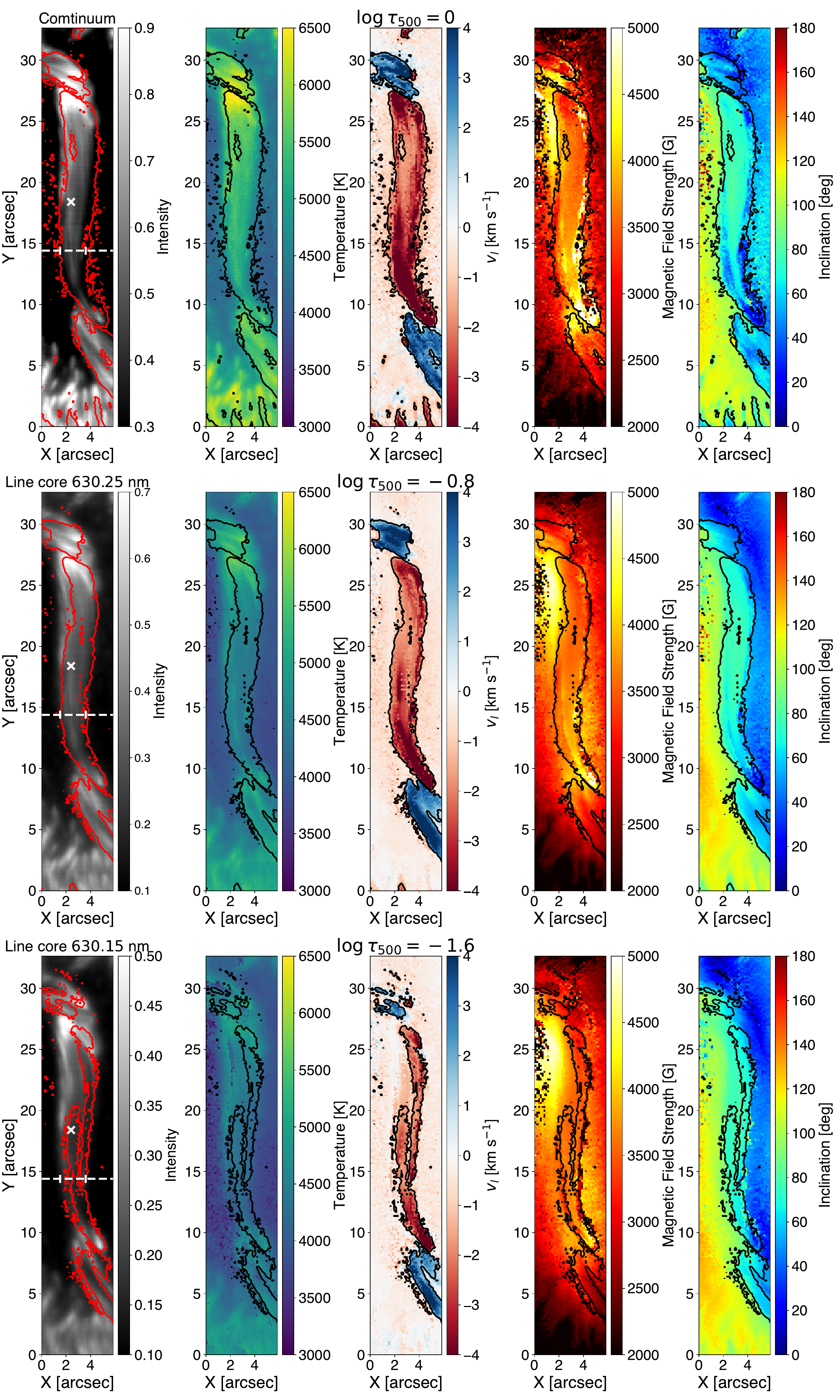}
  \caption{\jl{Hinode}/SP results for the \jl{central penumbral} light bridge for the 2017-09-06T 00:02:05 raster. From top to bottom (excluding leftmost column): maps at optical depth $\log \tau_{500} = 0$, $-0.8$ and $-1.6$. Leftmost column, from top to bottom: intensity maps for the continuum, line core at \jl{630.25 nm}, and line core at \jl{630.15 nm}. From left to right (from the second column): temperature, Doppler velocity (saturated at $\pm 4$~km~s$^{-1}$), magnetic field strength, and inclination. Zero degrees inclination is toward the observer. The contours are for $v_l=\pm 1.5$~km~s$^{-1}$. The cross in the upper-left panel marks the location of the example fitting results in Figure~\ref{fig:fitting}. The dashed line indicates the location of the vertical slice in Figure~\ref{fig:vertical}. The results are from the SIR model atmosphere with multiple nodes.}
  \label{fig:inversion}
  \vspace{2mm}
  \end{figure*}


\begin{figure}[t!]
\centering\includegraphics[width=0.45\textwidth]{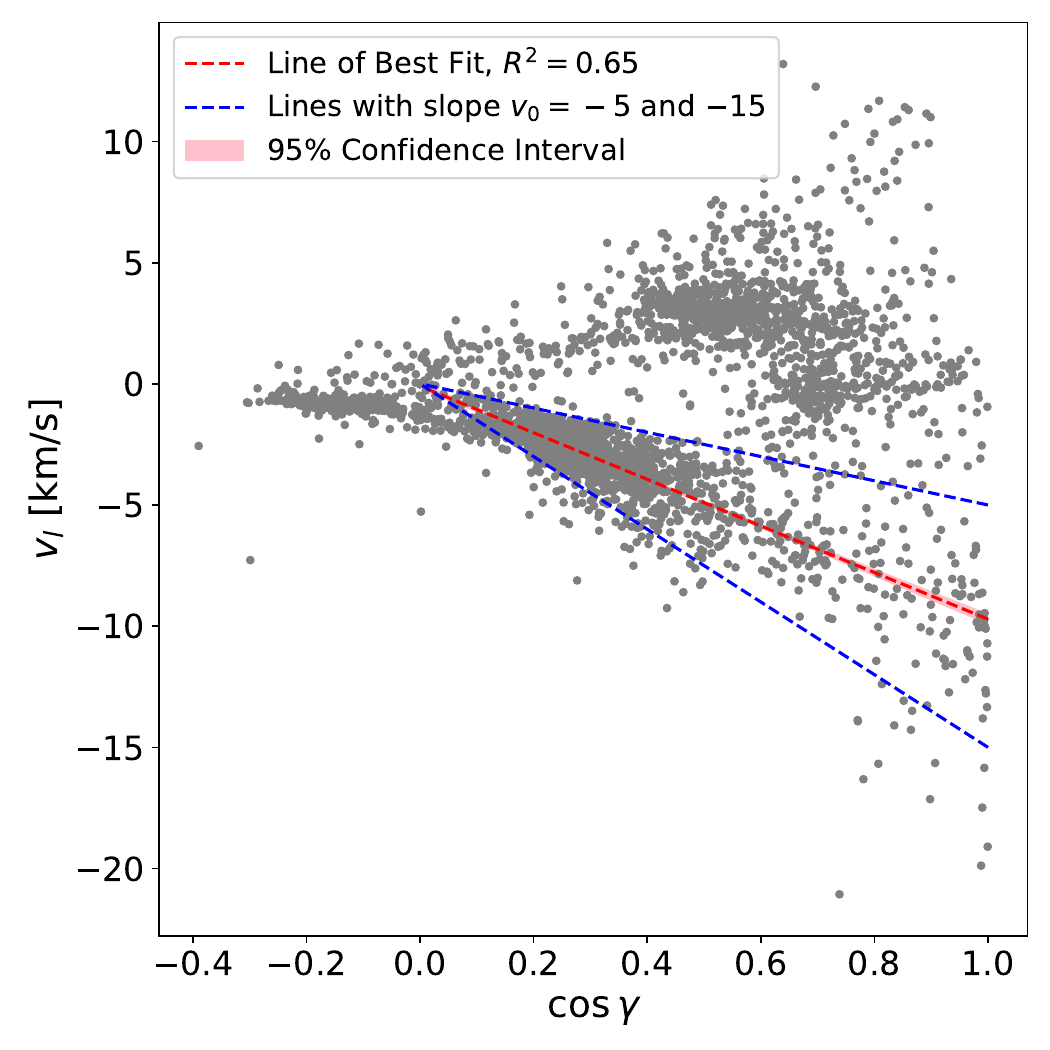}
\caption{Scatter plot of cosine of magnetic inclination $\cos \gamma$ and LOS velocity $v_l$ inside the light bridge, for pixels with continuum intensity greater than $30\%$ of nearby quiet Sun. Points above and below the $v_l=0$ line come from the blueshifted and redshifted regions, respectively. The blue dashed lines illustrate the expected patterns if the observed $v_l$ is due to a $5$~km~s$^{-1}$ and $15$~km~s$^{-1}$, field-aligned flow in the redshifted light bridge. The red line is the best fit for pixels with $v_l < -1.5$~km~s$^{-1}$. \jl{The pink shade illustrates the 95\% confidence interval, which is about twice the line thickness.} The slope is $9.6$~km~s$^{-1}$. The blueshifted points appear to lack a clear linear relation.}
\label{fig:scatter}
\vspace{2mm}
\end{figure}


\subsection{Fine-Scale and \jl{Depth-dependent} Structures}\label{subsec:sp}
\label{sec:sp}

We use \jl{Hinode}/SP rasters to study the fine-scale structures in the light bridge. Figure \ref{fig:inversion} shows the inferred atmospheric parameters at different optical depth. We also include the intensity maps at continuum and line cores at $630.25$ \jl{nm} and $630.15$ \jl{nm} in the leftmost column, which probe increasingly higher layers. At $0\farcs3$ spatial resolution, several penumbral filaments are discernible in the light bridge. Their north-south orientation is almost parallel to the PIL.

The second column of Figure ~\ref{fig:inversion} displays the inferred temperature, which decreases with height in the PIL area. The central, redshifted light bridge appears to be cooler than the blueshifted part of light bridge at all heights. There is clear intensity enhancement at line continuum and two line cores and temperature enhancement at the interface of the \jl{blueshifted} and \jl{redshifted} regions in the north ($Y\approx27\arcsec$) in all three heights. The intensity enhancement can also be observed on HMI observation too (Figure~\ref{fig:roi}). The highest temperature is at $\log \tau_{500} = 0$ and can reach about $6800$ K.

The third column of Figure~\ref{fig:inversion} shows the Doppler velocity. The pattern at $\log \tau_{500} = 0$ is overall consistent with the HMI results (cf. Figure~\ref{fig:roi}), though with a greater magnitude. The redshifted Doppler velocity increases from the center of the light bridge towards the northern and southern ends, and towards the western edge. The maximum reaches \jl{$-21.1$~km~s$^{-1}$}. We note that such high velocity will have adverse impact on the magnetic/velocity inference with default HMI data (see Section~\ref{subsec:flowtracking}). The transition from redshift to blueshift is quite drastic: $v_l$ changes from about $-4$ to $+3$~km~s$^{-1}$ over just a few pixels. The magnitude of Doppler velocity decreases in higher layers. The high Doppler velocity regions ($\pm1.5$~km~s$^{-1}$ contours) at $\log \tau_{500} = 0.0$ and $-0.8$ are nearly co-spatial with the light bridge. At $\log \tau_{500} = -1.6$, the high Doppler velocity regions are concentrated on the boundary between umbra and light bridge.

The fourth column of Figure~\ref{fig:inversion} shows the inferred magnetic field strength. At $\log\tau_{500}=0$, the light bridge has an overall stronger field strength compared to the surrounding umbrae. The southern end of the redshifted region has particularly strong field ($B>4500$~G), which is co-spatial with large Doppler velocity. The field strength decreases rapidly with height: the median decreases from $4950$~{G} at $\log \tau_{500} = 0$ to $3170$~{G} at $\log \tau_{500} = -1.6$. This is in contrast with the northern end, where the stronger field is located between the light bridge and the umbra to its east with low Doppler velocity.


The rightmost column of Figure~\ref{fig:inversion} shows the inferred magnetic inclination with respect to the LOS. It becomes more transverse (closer to $90\degr$) from $\log \tau_{500} = 0$ to $-1.6$: the median increases from \jl{$67.0\degr$ to $74.6\degr$}. A larger variation of inclination can be observed on $\log \tau_{500} = 0$. For the southern half of the central light bridge, the values decrease from about $90\degr$ in the center to about $20\degr$ at the southern end and the western edge. Larger Doppler velocity seems to appear where the magnetic field is more aligned with the LOS.


\begin{figure}[t!]
\centering\includegraphics[width=0.45\textwidth]{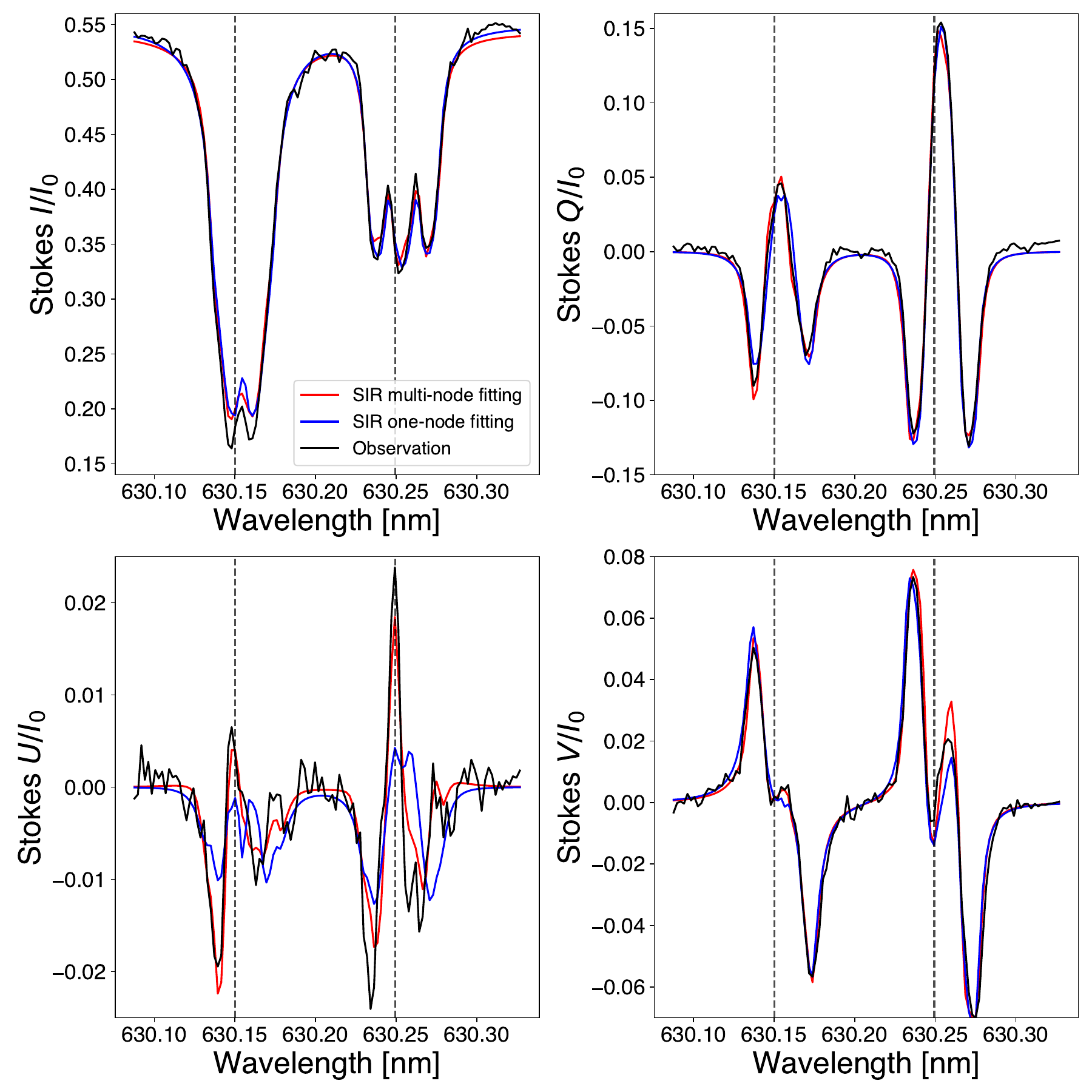}
\caption{Example of SIR fitting of \jl{Hinode}/SP observation. The four panels are for Stokes $I$, $Q$, $U$ and $V$, 
normalized by quiet-Sun continuum \jl{intensity} $I_0$. Black curves show the observed Stokes profile in the light bridge, for the point marked by the white $\times$ in Figure~\ref{fig:inversion}. Blue and red lines are for SIR fitting with one and multiple nodes, respectively. The two vertical dotted lines indicate the reference line-center wavelengths.}
\label{fig:fitting}
\vspace{2mm}
\end{figure}


In Figure~\ref{fig:scatter}, we show the scatter plot of the Doppler velocity $v_l$ and the cosine of the magnetic field inclination $\cos \gamma$ for pixels inside the light bridge. For the redshifted central portion, we observe a clear, quasilinear trend. This is in support of a largely field-aligned flow in the form $v_l=v_0 \cos\gamma$, with a constant flow speed $v_0$. The best fit line has slope $v_0 = -9.6 \pm 0.3$~km~s$^{-1}$ ($95\%$ confidence) and intercept of $-0.1$~km~s$^{-1}$. Two blue lines in Figure~\ref{fig:scatter} illustrate two expected field-aligned flow patterns with velocity $v_0 = $ $-5$~km~s$^{-1}$ and $-15$~km~s$^{-1}$. About $69\%$ of pixels in the central redshifted light bridge reside in this range.

We find that the multi-node inversion outperforms the single-node inside the light bridge, which implies gradients in the physical variables along the LOS. In Figure \ref{fig:fitting}, we show an example of the Stokes profiles for a typical pixel in light bridge, as well as the SIR fitted profiles with one and multiple nodes. The Stokes profiles exhibit complex shapes, with clear splitting in $I$, asymmetric red and blue lobes in $U$, and an additional lobe at around 630.26~nm for $V$. The one-node solution, which does not allow for gradient along the LOS, fits  the Stokes $U$ poorly,  in particular for the $\pi$ component. The multi-node model, on the other hand, markedly improves the fitting quality. Appendix~\ref{app:SP} presents a comparison of the quality of fits.

\begin{figure}[t!]
\centering\includegraphics[width=0.48\textwidth]{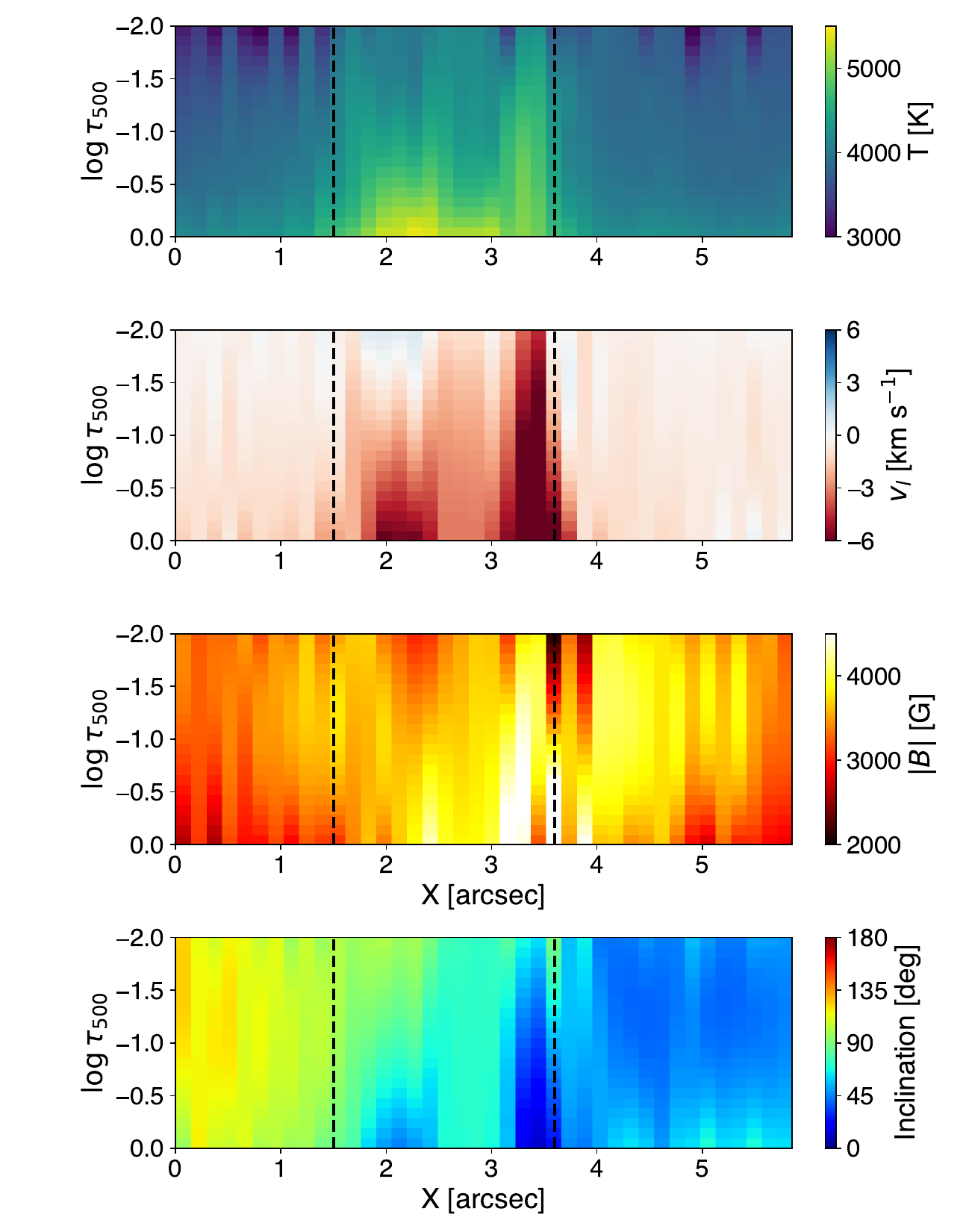}
\caption{A vertical slice through the light bridge along the dashed line in Figure~\ref{fig:inversion}. The $x$-axis shows the spatial extent. The $y$-axis shows the logarithm of optical depth, limited between $\log \tau_{500} = 0$ and $-2$. From top to bottom: temperature, Doppler velocity, and inclination with respect to the LOS. The black dashed lines indicate the approximate boundary of the light bridge, where the continuum intensity is greater than $30\%$ of the quiet Sun.} 
\label{fig:vertical}
\end{figure}

In Figure~\ref{fig:vertical}, we visualize the LOS variations using a vertical cut across the light bridge (dotted line in the left panel of Figure~\ref{fig:inversion}). This line crosses the high Doppler velocity region. The hottest region is near $X=2\arcsec$. The temperature reaches \jl{$5416$~K} at $\log\tau_{500}=0$, about $1000$ K higher than the surrounding umbrae. It then quickly decreases to \jl{$4335$~K} at $\log\tau_{500}=-1$. This coincides with a decrease of the Doppler velocity magnitude of \jl{$4.0$~km~s$^{-1}$} and an increase of the inclination of \jl{$30.7\degr$}. The magnetic field becomes more transverse to the LOS as height increases, and the Doppler velocity decreases. Figure~\ref{fig:vertical} additionally illustrates the filamentary sub-structure within the light bridge: at least two distinct strands are visible centered near $X=2\arcsec$ and $3.5\arcsec$.


\begin{figure*}[t!]
\centering\includegraphics[width=\textwidth]{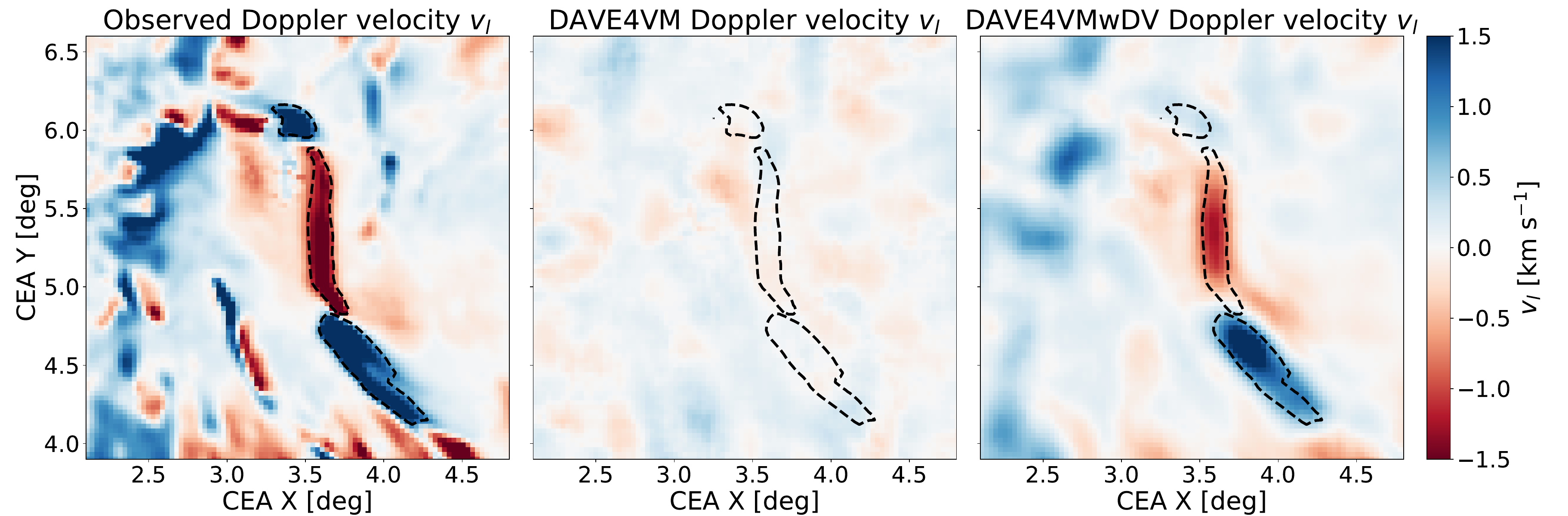}
\caption{Comparison of observed and inferred Doppler velocities \jl{from HMI}. Only the central region is shown. Left: averaged HMI $v_l$ map for 00:00 and 00:12 UT. Middle: $v_l$ from DAVE4VM. Right: $v_l$ from DAVE4VMwDV. The dashed contours are for observed $v_l=\pm1$~km~s$^{-1}$.} 
\label{fig:vl}
\end{figure*}


\begin{figure}[t!]
\vspace{3mm}
  \centering\includegraphics[width=0.48\textwidth]{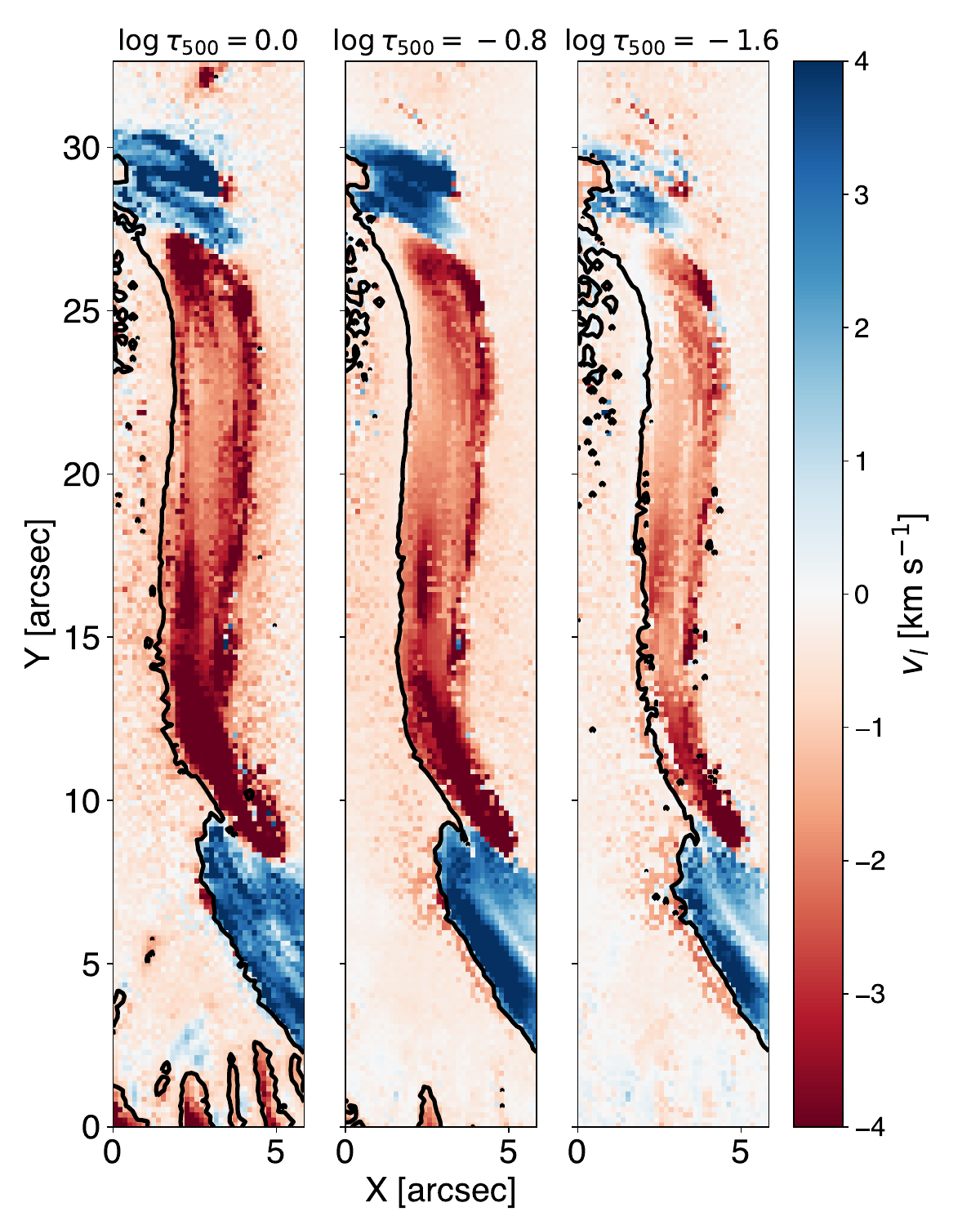}
  \caption{The Doppler velocity maps \jl{inferred from Hinode/SP observation} at optical depth $\log \tau_{500} = 0.0$, $\log \tau_{500} = -0.8$ and $\log \tau_{500} = -1.6$. The black lines are the PILs for the LOS magnetic field $B_l$ at each depth.} 
  \label{fig:PIL}
\end{figure}


\section{Discussion}\label{sec:Discussion}

\subsection{On the Flow Tracking Methods}\label{subsec:flowtracking}

The new Doppler constraint in the DAVE4VMwDV algorithm leads to results that are more consistent with the observed $v_l$. As shown in Figure~\ref{fig:vl}, the original DAVE4VM algorithm yields redshifts that are too low in the light bridge: the fast Doppler signals are completely missing. In comparison, they are partially recovered by DAVE4VMwDV. 

For the horizontal velocity field $\bm{v}_h$, both methods infer counterclockwise rotational flow patterns in the umbrae \citep[e.g.,][]{Yan2018}. DAVE4VMwDV additionally infers fast, converging flows inside the light bridge that are missing from the DAVE4VM result  (Figure~\ref{fig:flow}). For active regions with high Doppler velocity such as NOAA 12673, the different velocity fields are expected to yield significantly different energy and helicity flux estimates. 

We show in Appendix~\ref{app:DAVE4VMwDV} that the improved match with the observed $v_l$ (smaller $L_2$ in Equation set~(\ref{l2})) comes at the cost of a somewhat larger residual of the induction equation (greater $L_1$). We set $\lambda = 0.5$ in this study, which appears to \jl{strike} a reasonable balance between $L_1$ and $L_2$.

\jl{We consider three main sources of uncertainty in our inferred velocity:}

\jl{First, the ideal induction equation may not hold for a time sequence of magnetograms. For velocity inference, we make two assumptions of the input data: (1) a magnetogram presents $\bm{B}$ values from the same geometric height, and (2) the inferred $v_l$ values are also from that same height. Neither assumption is strictly true. For example, the Milne-Eddington inversion for HMI does not allow for LOS variations of the magnetic field. This results in effectively an ``averaged" $\bm{B}$ along the LOS, which does not necessarily correspond to any specific height. Previous studies also suggest that the inferred Doppler velocity and magnetic field are sensitive to different optical depths \citep{sensitivity}. It is thus not expected that the inferred $\bm{v}$ can closely reproduce the observed $v_l$ and adhere to the induction equation at the same time, even in the absence of noise. 
Additionally, the DAVE4VMwDV algorithm ignores the diffusion term in the induction equation. The effect of diffusivity is generally thought to be negligible, but can become important when intense current layers are present.}

\jl{Second, the input magnetograms contain both systematic and statistical uncertainty that can affect the downstream velocity estimates. The effect of certain systematics, such as the residual signal from the spacecraft velocity with respect to the Sun \citep{HMIinv}, has been investigated in detail \citep{schuck2016}. The statistical uncertainty induced by photon noise is expected to be small in the strong-field region, but may be amplified by the spatial and temporal derivatives. To quantify its impact, we generate a sample of magnetograms (size $N\sim 300$) with different noise realizations following the Monte Carlo method in \citet{avallone2020}, and subsequently infer a sample of velocity fields using DAVE4VMwDV. We adopt the sample's standard deviation (at each pixel) as our error estimate. We find that the the median relative errors are $(12\%, 14\%, 12\%)$ for the three velocity components $(v_x, v_y, v_z)$ in the region of large Doppler velocity ($|v_l| > 1$~km~s$^{-1}$), respectively. More details can be found in Appendix \ref{app:Mangetogram}.}

\jl{Third, specific numerical schemes may contribute to the errors. Here we evaluate two finite difference algorithms: (1) a five-points stencil method as described in section \ref{sec:D_M} and (2) the Scharr operator, on the same set of input magnetograms. We use the difference of their inferred velocity field to gauge the error. We find that the values from the two algorithms are highly correlated. The standard deviations are $(49.4, 53.5, 34.1,27.4)$~m~s$^{-1}$ for the differences of the inferred $(v_x, v_y, v_z, v_l)$, respectively. More details can be found in Appendix \ref{app:Derivative}. }


One issue specific to active region NOAA 12673 is its super fast flow velocity. \jl{The maximum reliable Doppler shift that can be detected by HMI is $\pm 7$ km s$^{-1}$, or an equivalent shift of $\Delta \lambda \simeq 144$~{m\AA} including the signal of satellite motion and solar rotation \citep{HMIshift},} the inferred $v_l=-21.1$~km~s$^{-1}$ (Section~\ref{subsec:sp}) that equivalent to a $432$~{m\AA} wavelength shift at $6173$~{\AA} will displace the \ion{Fe}{1} line completely outside the HMI spectra window. For velocity inference methods, the \jl{Courant-Friedrichs-Lewy} condition further stipulates an upper limit of $1$~km~s$^{-1}$ from HMI's $720$~km resolution and $720$~s cadence. The high Doppler shift, in violation of the \jl{Courant-Friedrichs-Lewy} condition, may induce artifact in the inferred velocity.

We note that the DAVE4VM(wDV) algorithms are fundamentally based on optimization: they do not require exact consistency between input and the observed $v_l$ and the induction equation. This may become an advantage when dealing with imperfect observations.


\begin{figure}[t!]
  \centering\includegraphics[width=0.4\textwidth]{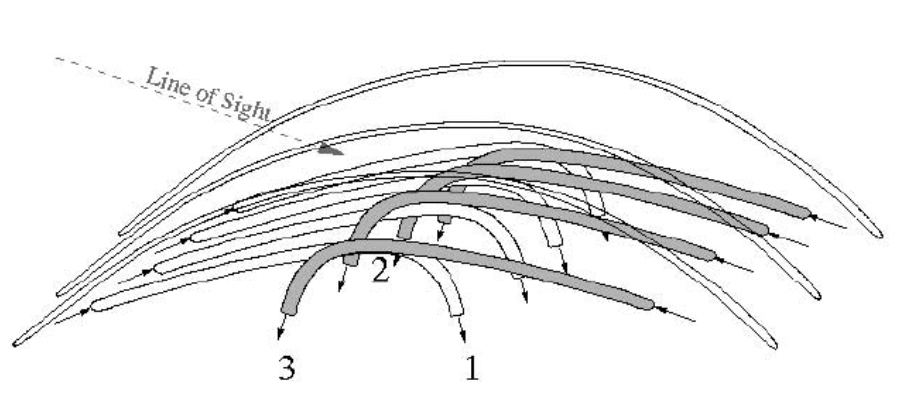}
  \caption{Scenario for explaining the persistent convergent flows near the PIL from \cite{fast_flow_Lite}. The magnetic field flux tubes interlace and bend downward near the PIL seen from LOS. The oppositely-directed Evershed flows converge upon the PIL and return to the interior.}
  \label{fig:Lite}
\end{figure}

\subsection{On the Origin of Large Photospheric Doppler Shift}\label{subsec:floworigin}

The penumbral light bridge in active region NOAA 12673 is in many ways similar to that in NOAA 11967 \citep{okamoto,Castellanos2020}. Both have strong, horizontal magnetic field; both show large Doppler velocity with alternating signs; the Doppler signals are both long-lasting. \cite{okamoto} suggested the fast flow was the projection of Evershed flow. The de-projected, field-aligned flow speed is $7.2$~km~s$^{-1}$, similar to $9.6$~km~s$^{-1}$ in our study. They proposed a scenario analogous to the crust subduction in the Earth's plate tectonics. The outflows originated from the two magnetic polarities colliding with each other; the stronger flow rolls under the weaker one. The flow then compresses the flux tubes and enhances the magnetic field in the light bridge.

One difference between these two active regions is the location of the PIL (of $B_l$) with respect to the Doppler patterns. \citet{Castellanos2020} showed that the redshifted and blueshifted flows are separated by the PIL at all heights in NOAA 11967 (see their Figure 3). They thus proposed a twisted, emerging flux rope as the origin of the LOS flows. We show in Figure \ref{fig:PIL} that the Doppler flows of both signs reside on the same side of the PIL in NOAA 12673. The flux rope scenario is thus less likely here. We note that flux emergence or submergence can still contribute, because the field-aligned flow only partially accounts for the observed Doppler velocity (Section~\ref{subsec:hmi}).

Fast converging/shearing flows along the PIL (Section~\ref{subsec:hmi}) were also found in other $\delta$-spots that have clear Doppler signals \citep[e.g.,][]{denker2007,prasad,cristaldi2014}. Their flow speeds are largely consistent with Evershed flows. In active region NOAA 12673, the flow is particularly fast near the southern end of the central light bridge where the Doppler velocity changes sign. This pattern is consistent with the picture proposed by \citet{fast_flow_Lite}: ``Evershed flows from opposite direction converge near the PIL, ...where the field lines bend downward and return their respective flows to the interior'' (Figure \ref{fig:Lite}). If the field lines are inclined enough and the active region is far away from the disk center, the two converging flows may produce Doppler signals of opposite sign due to projection.

While the Doppler signal in active region NOAA 12673 can be largely explained by field-aligned flow, the exact driving mechanism remains unclear. Numerical simulations suggested the important role of the Lorentz force, either for Evershed flow in an isolated sunspot penumbra \citep{rempel2011}, or for fast shear flow in a $\delta$-spot light bridge \citep{toriumi2019}. This is difficult to test based on our current analysis of the single \jl{Hinode}/SP observation  because the inversion was performed under the assumption of hydrostatic equilibrium. 

One other possibility is that gas pressure gradients drive these flows, i.e., they are siphon flows. Under the assumption total pressure balance, stronger gas pressure would equate with weaker magnetic pressure. The flow could be directed from the foot point of a flux tube with weaker magnetic field strength. Following the procedures in \citet{prasad2022}, we tested this hypothesis by comparing the field strength of the field line foot points in a nonlinear force-free field extrapolation model \citep{wiegelmann2012}. Unfortunately the result is inconclusive.  An alternative hypothesis is that factors such as differences in temperatures or heating / cooling rates might drive such flows.

We noticed enhancement in continuum intensity and temperature at the interface of redshifted and blueshifted flows at the northern end of the light bridge, i.e., around $(X,Y)=(2\arcsec,27\arcsec)$ in Figure~\ref{fig:inversion}. This could indicate interaction between the oppositely-directed flows or magnetic fields. For example, shocks may form and heat the local plasma because the estimated $9.6$~km~s$^{-1}$ flow speed likely exceeds the local sound speed \citep[about $7$~km~s$^{-1}$;][]{bellotrubio2003}.

\section{Conclusion}\label{sec:conclusion}

In this study, we investigate the origin of the large, persistent Doppler velocity near the PIL of active region NOAA 12673. We apply the new DAVE4VMwDV velocity estimator on \jl{SDO}/HMI data, and further analyze the fine-scale  structures in \jl{Hinode}/SP observation using the depth-dependent inversion algorithm SIR. Our main findings are as follows.
\begin{itemize}[parsep=0ex,partopsep=-0.5ex,itemsep=1ex,leftmargin=4mm]
   \item HMI observations show fast converging and shearing motions along the central PIL.
   \item The Doppler velocity signal can be mostly, but not entirely, explained by field-aligned flow in the $\delta$-spot light bridge. 
    \item SP observations suggest filamentary structures within the light bridge and gradients of physical variables along the LOS. 
    \item Doppler velocity and magnetic inclination inferred from SP data display a quasi-linear relation, consistent with a $9.6$~km~s$^{-1}$ field-aligned flow.
\end{itemize}

The fast Doppler flows near the PIL have been observed in other $\delta$-sunspots. Their properties are mostly consistent with the projected Evershed flow. In a follow-up work, we will further investigate the fine-scale structures of the light bridge by employing a two-component inversion scheme.

 
\begin{acknowledgments}
We thank Luis Bellot Rubio, Ivan Mili\'{c}, and Shin Toriumi for helpful discussions. J. Liu and X. Sun are supported by NSF award \#1848250 and the state of Hawai$\okina$i. 
B. Welsch gratefully acknowledges support from NASA LWS 80NSSC19K0072.
We thank the \jl{SDO}/HMI and \jl{Hinode}/SP teams for the data support. \jl{Hinode} is a Japanese mission developed and launched by ISAS/JAXA, collaborating with NAOJ as a domestic partner, NASA and STFC (UK) as international partners. Scientific operation of the \jl{Hinode} mission is conducted by the \jl{Hinode} science team organized at ISAS/JAXA. This team mainly consists of scientists from institutes in the partner countries. Support for the post-launch operation is provided by JAXA and NAOJ (Japan), STFC (U.K.), NASA, ESA, and NSC (Norway). The technical support and advanced computing resources from University of Hawai$\okina$i Information Technology Services – Cyberinfrastructure, funded in part by the NSF CC awards \#2201428 and \#2232862 are gratefully acknowledged.
\end{acknowledgments}



\appendix 


 \begin{figure}[t!]
  \centering\includegraphics[width=0.45\textwidth]{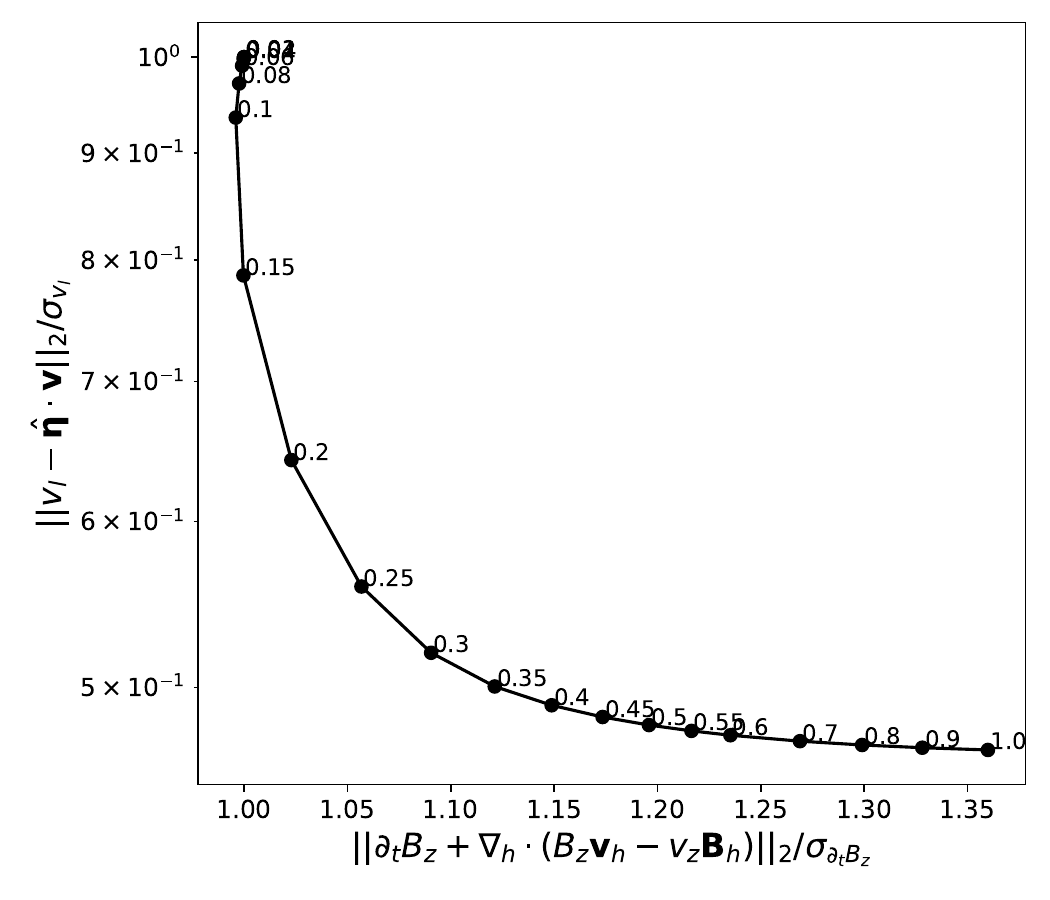}
  \caption{L-curve displaying the tradeoff between the two loss functions $L_1$ and $L_2$ in DAVE4VMwDV, for the HMI observations of active region NOAA12673. The values next to the black dots represent the weighting $\lambda$ of each test.}
  \label{fig:L_curve}
\end{figure}


\begin{figure}[t!]
  \centering\includegraphics[width=0.47\textwidth]{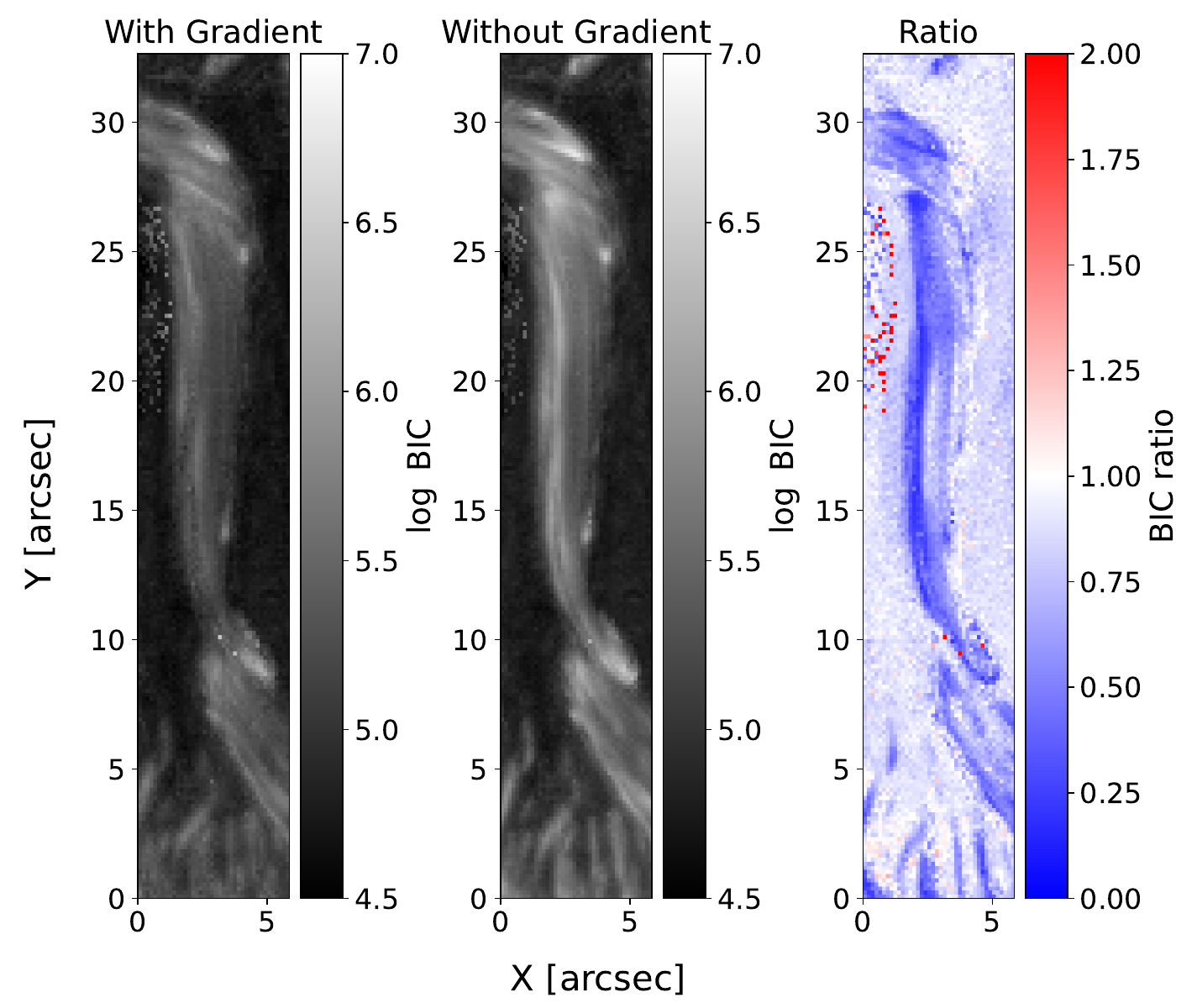}
  \caption{BIC maps of SIR fitting for the \jl{Hinode}/SP scan at 00:02:05 UT. Left: BIC map in logarithmic scale from inversion with multiple nodes (with gradient). Middle: same as left, but for single-node inversion (without gradient). Right: The ratio of BIC between multi-node and single-node inversions.}
  \label{fig:chi2}
  \vspace{2mm}
  \end{figure}


\section{Optimization of DAVE4VMwDV} \label{app:DAVE4VMwDV}

DAVE4VMwDV can be viewed as a regularized optimization algorithm for the residual of induction, i.e. term $L_1$ defined in Equation set~(\ref{l2}). The Doppler constraint $L_2$ serves as the regularization term. Besides the magnetogram and Dopplergram input, we need to specify the value of three free parameters: the window size $w$ for optimization, the maximum degree of Legendre polynomials $d$ for velocity expansion inside the window, and the relative weighting $\lambda$ for $L_2$.

We have experimented with several typical values of $w$ and $d$ previously used in DAVE4VM studies. In this study, we use $w = 23$ and $d = 7$. These empirical values provide a reasonable fitting for the Doppler velocity in active region NOAA 12673.

We vary $\lambda$ from $0$ to $1$ to test its impact on the final result. Figure~\ref{fig:L_curve} shows log-log curve of $(L_1,L_2)$, normalized by the respective uncertainties of the observational data. This "L-curve" (due to its "L" shape) displays the trade-off between the two terms as $\lambda$ varies \citep[e.g.,][]{Lcurve}. Here, $\lambda = 0$ represents the case of DAVE4VM, where we find a large discrepancy between the inferred and the observed $v_l$ (large $L_2$). As the weighting $\lambda$ increases from $0$ to $0.5$, the Doppler residual $L_2$ decreases drastically, while the induction term $L_1$ increases at a much slower pace. The opposite trend is true as $\lambda$ increases from $0.5$ to $1.0$. We finally select $\lambda=0.5$ as it offers a reasonable compromise: $L_2$ is reduced by a factor of $2$ compared to DAVE4VM, while $L_1$ only suffers a $20\%$ increase.

\begin{figure*}[t!]
  \centering\includegraphics[width=0.9\textwidth]{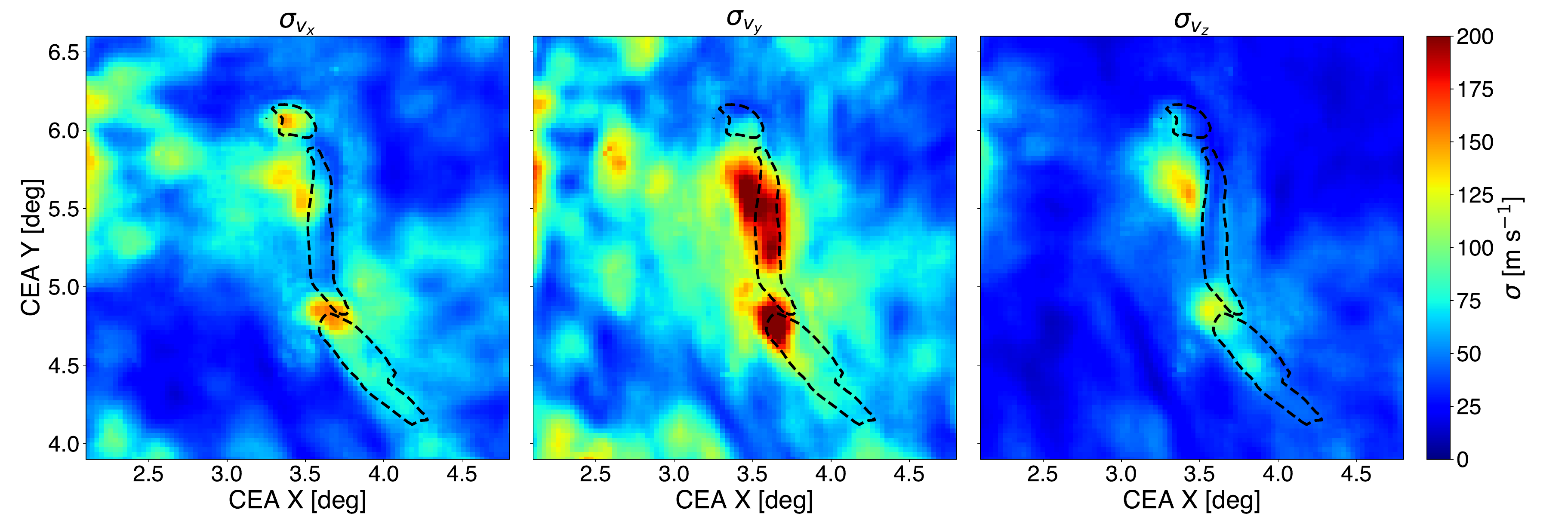}
  \caption{\jl{Maps of the standard deviations for $v_x$, $v_y$, and $v_z$ from the Monte Carlo sample, from left to right. Black dashed contours mark the region with $\lvert v_l \rvert > 1$ {km s$^{-1}$}.}}
  \label{fig:err}
    \vspace{2mm}
\end{figure*}


\section{\jl{Error Estimate}} \label{app:ERROR}

\jl{In Section \ref{sec:Discussion} we briefly discuss the three sources of errors in velocity estimate. Here we show the details of how the errors from HMI data and finite difference algorithms are estimated.}

\subsection{\jl{Error from Noise in HMI Data}}\label{app:Mangetogram}

\jl{We use a Monte Carlo method to estimate effect of noise in HMI data on the velocity estimate \citep{avallone2020}. The HMI pipeline provides formal error of the vector magnetograms in the form of variances and covariances of the magnetic field strength $B$, inclination $\gamma$, and azimuth $\psi$ variables. They are calculated based on the curvature matrix and the $\chi^2$ of the least-square-based inversion algorithm \citep{HMIshift}.}

\jl{Assuming the uncertainties are normally distributed, we create a correlated random sample of $(B,\gamma,\psi)$ with size $N\sim 300$ using the provided variances and covariances. We subsequently decompose the magnetic field vectors in to $(B_x,B_y,B_z)$ components and project the maps into the CEA coordinate \citep{sun2013arXiv}. These procedures yield $N$ magnetograms for each time step with different noise realizations, which are then used as the input of DAVE4VM for velocity estimate. Finally, we take the standard deviation of the $N$ velocity estimates at each pixel as the error caused by the noise of input.}

\jl{Figure \ref{fig:err} shows the maps of these standard deviations. The majority of pixels have values below $100$~m~s$^{-1}$; the median values are $(82.5,108.8,61.6)$~m~s$^{-1}$ for the three velocity components $(v_x, v_y, v_z)$, respectively. The absolute errors appear to be greater for $v_y$, especially around the light bridge with high shear/converging velocity (c.f. Figure~\ref{fig:flow}). The relative errors are similar for all components, on the order of $10\%$.}
 


\subsection{\jl{Error from Finite Difference Algorithms}}\label{app:Derivative}

\jl{To estimate the error introduced by finite difference algorithms, we use the five-points stencil and the Scharr operator methods and compare the inferred velocities. The five-points stencil method is the default algorithm in DAVE4VM and is described in Section \ref{sec:D_M}. The Scharr operator is a filtering method that is often used to find the spatial derivative of an image. It is defined as}
\jl{\begin{equation*}
    h^\prime_x = 
    \begin{pmatrix}
     47 & 0 & -47 \\
     162 & 0 & -162 \\
     47 & 0 & -47 \\
    \end{pmatrix},
  \qquad 
  h^\prime_y = 
  \begin{pmatrix}
   47 & 162 & 47 \\
   0 & 0 & 0 \\
   -47 & -162 & -47 \\
  \end{pmatrix}
\end{equation*}}
\jl{for the $x$ and $y$ direction, respectively. The derivatives can be calculated by convolving (operator $\ast$) the 2D array with the Scharr operator, i.e., 
\begin{equation}
  \begin{split}
  \frac{\partial f_{i,j}}{\partial x} &= (f \ast h^\prime_x)_{i,j}, \\
  \frac{\partial f_{i,j}}{\partial y } &=(f \ast h^\prime_y)_{i,j}.
  \label{diff}
  \end{split}
\end{equation}
}

\begin{figure}[t!]
  \centering\includegraphics[width=0.48\textwidth]{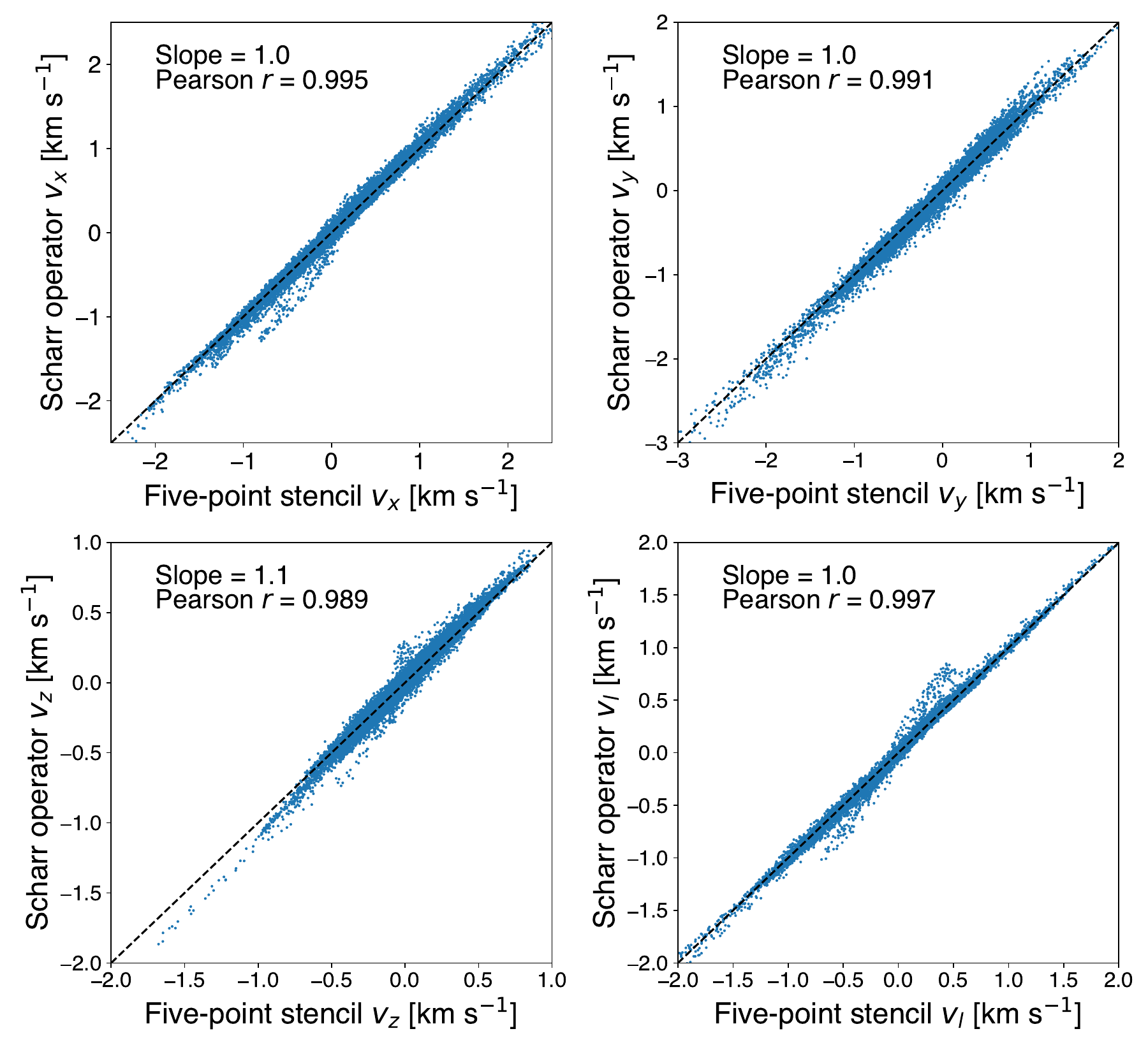}
  \caption{\jl{Scatter plots of the velocities inferred using five-point stencil and the Schall operator for numerical differentiation. The four panels are for the $v_x$, $v_y$, $v_z$, and $v_l$ component, respectively. The slope of a linear fit and the Pearson correlation coefficient $r$ are noted for each case. }}
  \label{fig:err_diff}
\end{figure}

\jl{The scatter plots of the inferred $v_x$, $v_y$, $v_z$, and $v_l$ based on these two algorithms are shown in Figure \ref{fig:err_diff}. The slopes are all close to $1$ and have Pearson correlation coefficient of about $0.99$. The scatter is largest in $v_z$, suggesting that its quantity is the most affected. }

\section{Hinode/SP Inversion} \label{app:SP}

In this study, we invert the observed Stokes profiles using the configuration in Table~\ref{tab:SIR} with twelve possible initial-guess atmosphere models derived by applying perturbations to the FAL atmosphere models \citep{Fontenla}. We pick the best fit by using the fitting with the largest coefficient of determination, $R^2$, of Stokes $I$. 

In Section \ref{sec:sp} we described the comparison between inversion with one node (no gradients along LOS) and multiple nodes for a single pixel. To assess the overall performance, we calculated the Bayesian Information Criterion \citep[BIC;][]{BIC} following \cite{model_selection}. The BIC is defined as 
\begin{equation}
    \mathrm{BIC} = \chi^2 + k\ln N,
\end{equation}
where 
\begin{equation}
    \chi^2 = \sum_{n=1}^4\sum_{i=1}^N \, w_n^2 \,\left(\frac{I^\mathrm{obs}_n(\lambda_i)-I^\mathrm{syn}_n(\lambda_i)}{\sigma_i}\right)^2
\end{equation}
characterizes the differences between the observed (superscript ``$\mathrm{obs}$'') and modeled (``$\mathrm{syn}$'') Stokes parameters $I_i$ (normalized by uncertainty $\sigma_i$). It also combines the four Stokes parameters $(I_1,I_2,I_3,I_4)=(I, Q, U, V)$ with empirical weightings $(w_1,w_2,w_3,w_4)=(2, 5, 5, 5)$.  The number of free parameters for the single-node inversion is $k=10$; for the multi-node inversion there is $k=17$. The number of observations is $N=448$. The inversion with the smaller BIC is the preferred one.

Figure~\ref{fig:chi2} displays the BIC maps of the multi-node and single-node inversion, as well as the ratio between them. The ratio is mostly below $1.0$ in the light bridge, suggesting that the multi-node inversion result is preferred.


\bibliographystyle{aasjournal}
\bibliography{reference}

\end{CJK*}

\end{document}